# Resolving Capillary Mode Transitions in Microparticles at Fluid Interfaces

Sungwan Park,[1] Justin Jeongwoo Choi,[1] and Albert Tianxiang Liu[1,2,3,4*]

[1]*Department of Chemical Engineering, University of Michigan, Ann Arbor, MI, 48109, USA*
[2]*Department of Material Science and Engineering, University of Michigan, Ann Arbor, MI, 48109, USA*
[3]*Macromolecular Science and Engineering, University of Michigan, Ann Arbor, MI, 48109, USA*
[4]*Biointerfaces Institute, University of Michigan, Ann Arbor, MI, 48109, USA*

Capillarity-driven self-assembly at fluidic interfaces offers a scalable route to large-scale, reconfigurable materials. Microscale particles with high horizontal–to–vertical aspect ratios have emerged as attractive building blocks for controlled two-dimensional self-assembly thanks to the well-defined interfacial deformations they induce. Nevertheless, the underlying dynamics governing lateral capillary interactions at this intermediate length-scale remain incompletely understood. Here, we combine experiments and theory to elucidate the interplay among gravitational forces, wetting, and contact line undulations that give rise to curvature-mediated interparticle potentials. We focus on the transition between two capillarity regimes, monopolar and quadrupolar, using sub-millimeter, disc-shaped micro-particles at fluidic interfaces. By systematically reducing the lateral dimensions of the micro-discs, we probe the limits of the Bond number (Bo) in predicting this transition. Our results reveal that Bo alone is insufficient, as it omits key material parameters such as particle density and surface topography. We identify and experimentally validate a coupled set of parameters that, below a critical threshold, reliably predict the emergence of quadrupolar capillarity, driven by topography-induced meniscus undulations. We further derive a closed-form transition criterion that couples particle size, density, and quadrupolar amplitude through an effective drag term that accounts for deformation-induced dissipation, yielding quantitative agreement with simulations and experiments across particle pairings and interfaces. This study provides the first direct experimental demonstration of the monopolar–to–quadrupolar transition in a single particle system without compositional changes. The resulting model maps the governing parameter space and enables the controlled assembly of hierarchical two-dimensional super-structures with tunable phase behavior, offering design principles for next-generation interfacial materials built from increasingly miniaturized particulate building blocks.

## I. INTRODUCTION.

The self-assembly of microparticles at fluid–fluid interfaces provides a robust route to higher-order materials with broad technological relevance, including surface coatings [1,2], photonic systems [3–6], drug delivery [7,8], and microrobotics [9–11]. Adsorbed particles deform the interface, creating menisci that mediate lateral capillary interactions and drive spontaneous organization [12,13]. These interactions reflect a coupled balance of gravitational forces [14,15], interfacial tension [16], and particle-specific properties, including geometry [17–21], surface topography [22,23] and chemistry [24–28]. Their relative significance shifts with particle size, giving rise to distinct interaction regimes.

At larger scales, gravity dominates, producing **capillary monopoles**: radially symmetric interface deformations with no angular dependence [29]. Neighboring particles with same sign monopolar distortions (i.e., both curving up or down) generate overlapping menisci that reduce the interfacial free energy [14,15] and yield attractive interactions [25]. Conversely, unlike monopolar menisci with opposite curvature fields produce capillary repulsion [15,30,31]. This behavior is often compared to electrostatics, with interfacial rises and depressions described as positive and negative "capillary charges" [12,29]; however, unlike electrostatics, same-signed capillary charges attract.

As particle size decreases toward the colloidal regime, gravitational forcing becomes negligible and monopolar deformations vanish [13,32]. Lateral capillarity is instead governed by **capillary quadrupoles** arising from contact line undulations induced by surface roughness [22]. Even nanometric roughness can produce a puckered contact line [22,33,34], often represented as a superposition of multipolar modes (e.g., quadrupole, hexapole, octapole, etc.) [35,36]. Higher-order modes decay rapidly, so quadrupoles dominate long-range interactions at small scales [18,22]. A defining feature of quadrupolar capillarity is its universally attractive nature: due to their alternating capillary charges with a two-fold axial symmetry, quadrupolar particles can always rotate into configurations that enable attractive interactions, provided they retain rotational degrees of freedom [13].

The Bond number ($Bo$) is the classical dimensionless group used to assess the relative contributions of gravity and surface tension in particle-laden fluidic interfaces:

$$Bo = \Delta\rho g {r_p}^2/\gamma \tag{1}$$

*Contact author: atliu@umich.edu

where $\Delta\rho$ is the density difference between the two fluids, $g$ is gravitational acceleration, $r_p$ is the particle radius, and $\gamma$ is the interfacial tension. When $Bo \ll 1$ (e.g., $Bo{\sim}10^{-5}$), interfacial tension is expected to overwhelm the gravitational forces exerted by the particle on the interface, and monopolar interactions are often presumed negligible [12,13].

However, quadrupole-dominated interactions have been reported at substantially larger $Bo$ (Fig. 1a), indicating $Bo$ alone may not be a sufficient predictor of the governing capillary mode. This limitation is anticipated because $Bo$ omits key determinants of quadrupolar distortions, such as particle surface topography, and does not explicitly capture how particle density and wettability set the magnitude and sign of monopolar curvature. Moreover, a pronounced experimental gap exists for particles in the transitional size range ($r_p \sim 10^{-4}$–$10^{-3}$ m, Fig. 1a), leaving the monopole–quadrupole boundary poorly defined.

Herein, we present direct experimental evidence resolving the monopole–quadrupole transition in a compositionally invariant disc-shaped particle system. By varying size and wettability at oil/water interfaces, we identify a predictive crossover threshold and develop a quantitative model that agrees with empirical measurements. We demonstrate that a system of microparticles that repel due to opposing monopolar distortions can transition to attractive behavior as size decreases and quadrupolar interactions dominate. These findings clarify how particle size, wettability, and topography jointly govern capillary forces, providing a framework for designing programmable interfacial self-assembly at fluidic interfaces.

## II. RESULTS and DISCUSSION

Micro-disc interfacial deformation falls into one of two distinct cases based on particle density. In Case 1 (e.g., light oil/water), particles denser than both fluids produce downward menisci regardless of wettability. In Case 2 (e.g., water/heavy oil), particle density lies between phases and monopolar curvature becomes wettability-dependent [25].

We fabricated hydrophilic PEGDA (1.130 ± 0.012 g/mL) and hydrophobic TMPTA (1.013 ± 0.032 g/mL) micro-discs via degassed micromolding lithography (Appendix A, Fig. 1b). Contact angles confirmed their contrasting wettabilities (Fig. 1c, Fig. S2, Table S1). Large discs ($r_p = 625$ μm) exhibit clear monopolar curvature (Fig. 1d). In Case 1, both PEGDA and TMPTA generate downward deformations, yielding negative monopolar charges (Fig. 1e) and mutual attraction despite opposing wettabilities (Fig. 1h). In Case 2, wettability dictates charge polarity: hydrophilic PEGDA produce negative monopoles, repelling positive hydrophobic TMPTA (Fig. 1e, bottom, Fig. 1i).

At smaller sizes ($r_p = 125$ μm), monopolar deformations vanish (Fig. 1f). Nanoscale surface roughness (AFM, Fig. S3) induces contact-line undulations that decay into a dominant quadrupolar mode [22,37,38]. Interferometry shows localized perimeter distortions; near-field profiles reveal a puckered contact line, while far-field profiles ($r = 2r_p$) exhibit two-fold symmetry, confirming quadrupolar dominance (Fig. 1g). Extracted quadrupolar amplitudes $H_2$ at the light oil/water interface are $1.10 \pm 0.18$ μm (PEGDA) and $1.21 \pm 0.11$ μm (TMPTA). Between water/heavy oil, they become $1.08 \pm 0.45$ and $4.20 \pm 2.86$ μm, respectively, consistent with prior reports [10]. Crucially, these quadrupolar discs attract robustly at both interfaces (Fig. 1j,k), in stark contrast to their monopolar counterparts (Fig. 1h,i).

Gel trapping (GTT, Figure S3) preserved equilibrium configurations (Fig. 1l,m), confirming pinned contact lines [37]. Together, these data demonstrate that monopolar discs attract in Case 1 but repel in Case 2 when charges oppose, whereas quadrupolar discs attract universally. This in turn also provides the first direct evidence of attractive quadrupolar interactions between particles completely residing in different fluidic phases.

To probe the monopolar–quadrupolar transition, PEGDA (hydrophilic) and TMPTA (hydrophobic) micro-discs of five radii (125–375 μm; fixed thickness $T$ = 85 μm) [39] in both Case 1 and 2 were tracked over time for PEGDA–PEGDA, TMPTA–TMPTA, and PEGTA–TMPTA pairs until motion ceased ($t = t_f$). Normalized center-to-center distance $L/2r_p$ versus remaining time $t_f - t$ (Fig. 2a–j, representative videos: Movie S1) exhibit a classic power-law dependence [17]:

$$\frac{L}{2r_p} \sim \left(t_f - t\right)^{\alpha} \qquad (2)$$

where $\alpha$ reflects the dominant interaction mode. In the range $r_p < L < \lambda_c$ (capillary length $\lambda_c = \sqrt{\gamma/\Delta\rho g}$; measured $\lambda_c = 4.54$ mm for light oil/water, 2.37 mm for water/heavy oil [29,35]), monopolar and quadrupolar forces scale as $F_{\mathrm{cap},0} \sim L^{-1}$ [29,40] and $F_{\mathrm{cap},2} \sim L^{-5}$ [13,26], respectively. In the overdamped, low-Reynolds regime, particle velocity $v = dL/dt \propto F_{\mathrm{cap}}$. Writing $F_{\mathrm{cap}} \sim L^{\beta-1}$ gives $\log\, t \sim (2-\beta)\log\, L$ [41], and combining with Eq. (2) yields

*Contact author: atliu@umich.edu

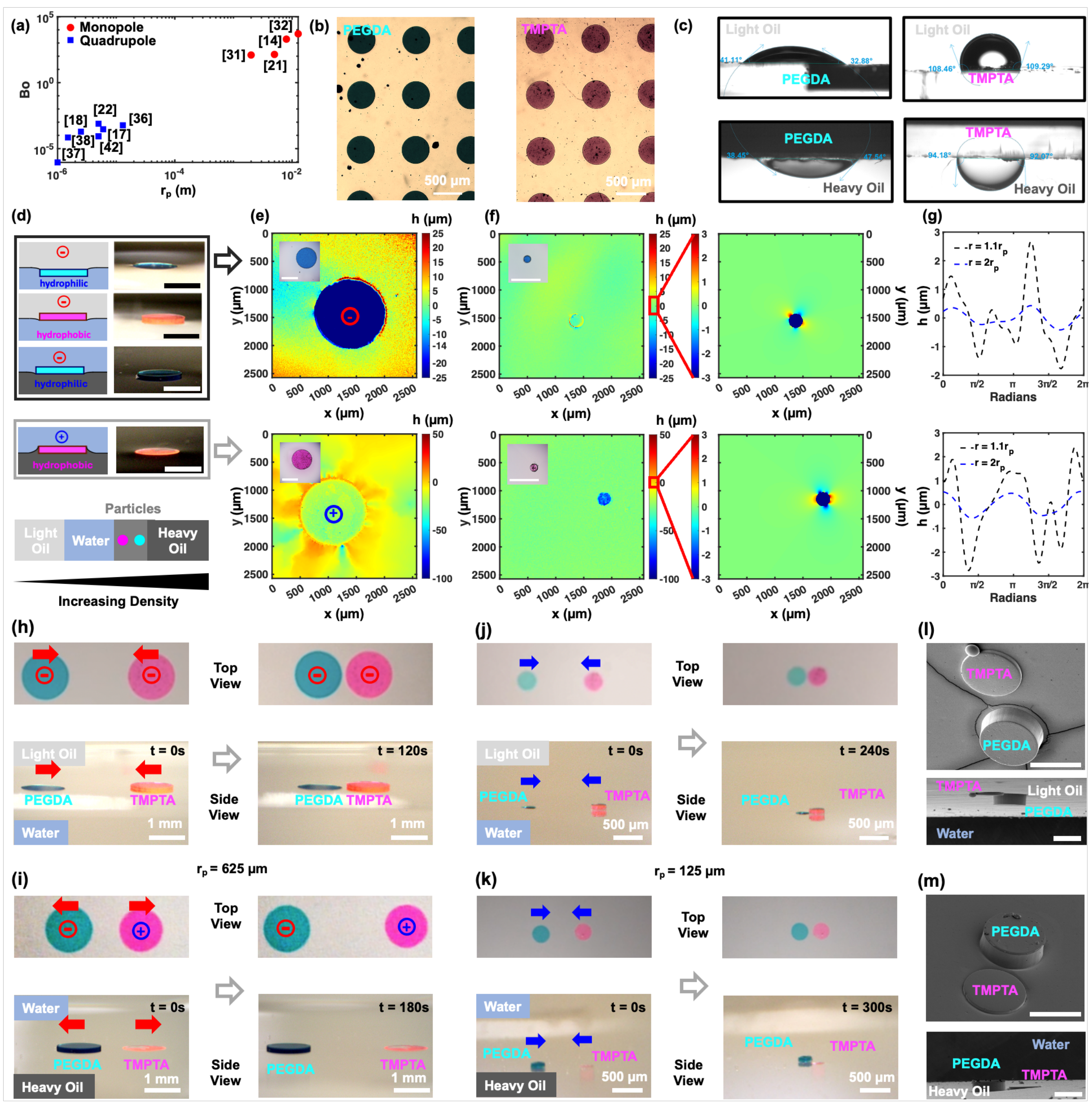

**Figure 1. Capillary interactions of disc-shaped particles at fluid interfaces.** (a) Bond number $Bo$ vs. particle radius $r_p$ for reported monopolar (red circles) and quadrupolar (blue squares) interactions; the absence of data between regimes highlights the unexplored transition range. (b) PEGDA (left) and TMPTA (right) micro-discs ($r_p = 250$ µm) in PDMS mold. (c) Water contact angles on PEGDA and TMPTA films in light oil (top) and heavy oil (bottom). (d) Schematics (left) and optical images (right) of large discs ($r_p = 625$ µm) at both interfaces; red/blue borders denote negative/positive capillary charge. Scale bars: 1 mm. (e) Interferometry height maps showing negative (top) and positive (bottom) monopolar charges; insets: bright-field images. Scale bars: 1mm. (f) Interferometry maps (left) and processed surface profiles (right) for smaller micro-discs ($r_p = 125$ µm) at the same interfaces, highlighting the absence of long-range monopolar curvature and the emergence of near-perimeter quadrupolar distortions; insets: bright-field images. Scale bars: 500 µm. (g) Interfacial height profiles from (f) at radial distance $r = 1.1r_p$ (black dashed) and $2r_p$ (blue dashed), revealing quadrupolar features. (h–k) Time-lapse top and side views of capillary interactions between a PEGDA and a TMPTA micro-disc: (h) monopolar discs at the light oil/water interface (Case 1) attract due to matching negative menisci; (i) monopolar discs at the water/heavy oil interface (Case 2) repel due to opposite capillary charges; (j) quadrupolar micro-discs at the light oil/water interface (Case 1) show mutual attraction;

*Contact author: atliu@umich.edu

(k) quadrupolar micro-discs at the water/heavy oil interface (Case 2) attract despite opposite wettability. (l,m) SEM images of PEGDA and TMPTA discs at (l) light oil/water and (m) water/heavy oil interfaces via gel trapping (GTT). Scale bars: 200 μm.

the exponent relation $\alpha = 1/(2-\beta)$ [17], with $\alpha = 1/2$ and $\alpha = 1/6$ correspond to monopolar and quadrupolar interactions, respectively.

Trajectory scaling reveals the capillary regime transition. At the light oil/water interface (Case 1, Fig. 2a–e), smallest discs ($r_p = 125$ μm) exhibit $\alpha \approx 1/6$, characteristic of quadrupolar attraction. As disc size increases, $\alpha$ shifts toward $1/2$, signaling the onset of monopole-dominated interactions. Both PEGDA and TMPTA carry negative capillary charges in Case 1; thus, all pairs attract regardless of regime transition.

At the water/heavy oil interface (Case 2, Fig. 2f–j), behavior is qualitatively different. Here, monopolar charge polarity depends on particle wettability. For $r_p \leq 250$ μm, all pairings remain quadrupolar ($\alpha \approx 1/6$). At $r_p = 312.5$ μm, unlike pairs (PEGDA–TMPTA) exhibit bifurcated behavior: 10 of 20 pairs attracted and 10 repelled, indicating competitive monopolar repulsion and quadrupolar attraction. At $r_p = 375$ μm, like pairs follow monopolar scaling ($\alpha \approx 1/2$), while unlike pairs predominantly repel (Movie S3), unambiguously marking the monopolar transition.

Near particle contact ($L/2r_p > L_f/2r_p + 0.3$), all trajectories stall and deviate from ideal power laws due to puckered contact lines that frustrate near-field alignment [36,42,43], leaving finite gaps between trapped particles (Fig. 1l,m). To further quantify the transition, we extracted $\alpha$ via $\log L = b + \alpha \log(t_f - t)$ over $L_f + 0.6r_p < L < \lambda_c$ [44]. Figure 3a,b plots $\alpha$ versus $r_p$ for all pair types. In both systems, $\alpha$ rises monotonically from $\approx 1/6$ to $\approx 1/2$ with increasing size. For Case 2, $\alpha$ values for unlike pairs at $r_p \geq 312.5$ μm are not reported, because strong monopolar repulsion precludes such fits.

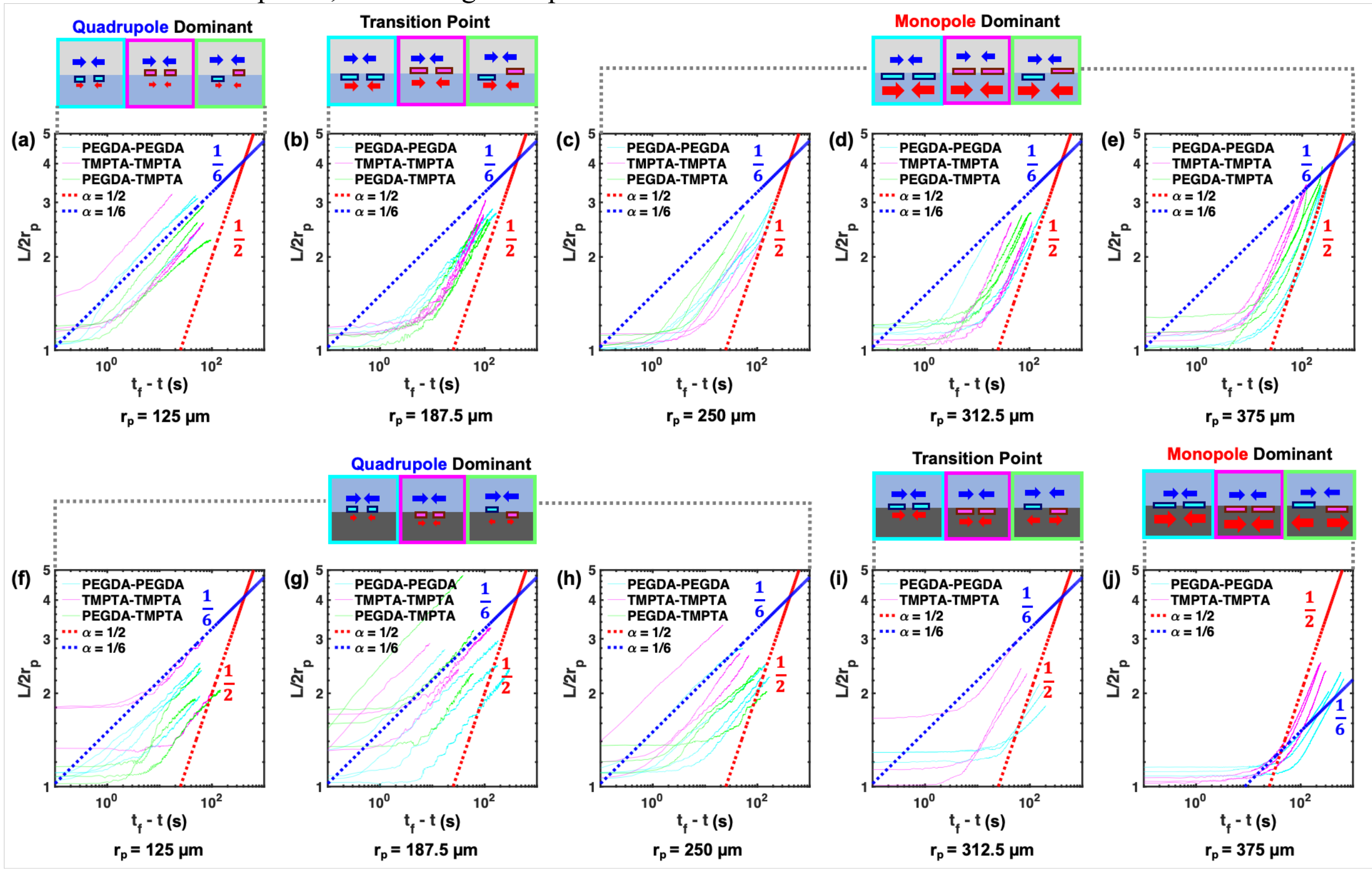

**Figure 2. Trajectory scaling of micro-disc pairs across interfaces and size.** (a–e) Normalized trajectories versus remaining time $(t_f - t)$ for the light oil/water interface (Case 1) for particles with radii of (a) 125, (b) 187.5, (c) 250, (d) 312.5, and (e) 375 μm, respectively. (f–j) Corresponding trajectories at the water/heavy oil interface (Case 2) for particles with the same radii. Dashed power-law guides with exponents of $\frac{1}{2}$ (red) and $\frac{1}{6}$ (blue) indicate the expected scaling for quadrupolar- and monopolar-dominated interactions, respectively. Schematic illustrations above the panels summarize the dominant interfacial deformation and interaction mechanism.

*Contact author: atliu@umich.edu

We model the capillary regime transition by explicitly incorporating (i) particle density $\rho_p$, which helps set the monopolar amplitude $H_0$, and (ii) surface topography that controls the quadrupolar amplitude $H_2$ (Fig. 3c–d). For capillary monopoles, the meniscus height $H$ at the contact line satisfies [40,45–47]:

$$H = \sin(\psi)\, r_p \ln\left(\frac{2\lambda_c}{\gamma r_p}\right) \quad (3)$$

where $\psi$ is the meniscus contact angle [32,48]. $H$ and $\psi$ are obtained from vertical force balance. For a PEGDA disc at the light oil/water interface (Case 1) [49]:

$$2\pi r_p \gamma \sin(\psi) + \pi r_p^2 (H+T) g \rho_I - \pi r_p^2 T g \rho_p = 0 \quad (4)$$

where $\rho_I, \rho_{II}$ are subphase/superphase densities, and $T$ is micro-disc thickness. The sign of the surface tension term reflects monopolar charge polarity (e.g., negative for all particles at light oil/water, giving a positive term in Eq. 4). Solving Eqs. 3 and 4 yields $H$ and $\psi$, defining the monopolar amplitude $H_{0,i} = \pm H$ [29,35]. Analogous expressions for the other three particle–interface combinations are derived in the SI (Equations S7 and S8, Section 6).

The total lateral capillary interaction force $F_{cap}$ is modeled as the sum of monopolar ($F_{cap,0}$) and quadrupolar ($F_{cap,2}$) contributions: $F_{cap} = F_{cap,0} + F_{cap,2}$, where the monopole–monopole capillary interaction force ($F_{cap,0}$) is:

$$F_{cap,0} = -2\pi\gamma H_{0,1} H_{0,2}/L \quad (5)$$

and the quadrupole–quadrupole capillary interaction force ($F_{cap,2}$) is:

$$F_{cap,2} = -48\pi\gamma H_{2,1} H_{2,2} r_p^4 / L^5 \quad (6)$$

where $H_{2,i}$ (i = 1,2) is the quadrupolar amplitude of particle $i$ [26], measured experimentally via optical profilometry [10,36,50]. Together, Equations (5) and (6) predict dominant capillary regime based on particle parameters (size, density, surface roughness) and separation distance $L$.

We simulated pairwise trajectories across particle radii (10–500 µm) and separations $L = 2r_p$ to $6r_p$. Simulated force profiles (Movie S4) reproduce the purely attractive interactions at the light oil/water interface across all sizes and capture the emergence of monopolar repulsion at the water/heavy oil interface with increasing size. Extracted trajectory exponents $\alpha$ (via $\alpha = 1/(2-\beta)$ from $\log(-F_{\mathrm{cap}}) \sim (\beta-1)\log L$ [17,44]) agree quantitatively with experiments across all pair types and interfaces (Fig. 3a,b, dashed curves).

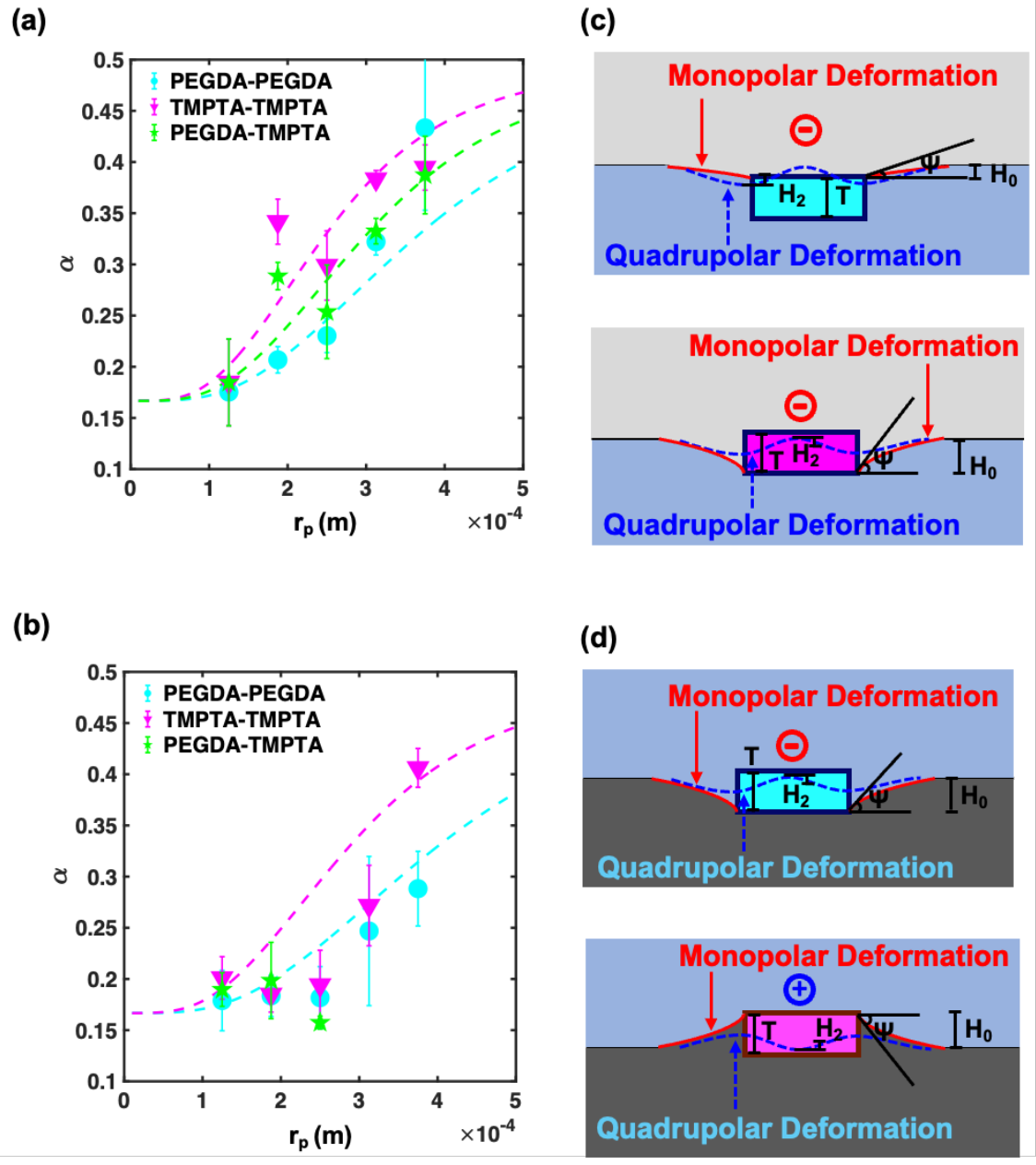

**Figure 3. Extracted power-law exponents reveal a size-dependent transition in capillary interaction mode.** (a–b) Fitted $\alpha$ values for PEGDA–PEGDA, TMPTA–TMPTA, and PEGDA–TMPTA micro-disc pairs at the (a) light oil/water and (b) water/heavy oil interfaces, plotted as a function of particle radius. Data points with error bars represent fitted values; dashed curves denote analytical-model predictions. (c–d) Schematic illustrations of combined monopolar (solid red) and quadrupolar (dashed blue) interfacial deformations for representative cases with positive or negative monopolar capillary charge.

Beyond reproducing the measured trajectories, the model allows us to simultaneously sweep multiple independent parameters (i.e., particle radius (via $Bo$), density $\rho_p$, and quadrupolar amplitude $H_2$) to map the crossover region. Figures 4a and 4e show three-dimensional (3D) surface plots of model-predicted $\alpha$ values for PEGDA–PEGDA and TMPTA micro-disc pairs in Case 2. At fixed surface roughness (e.g., $H_2 \sim 3$ µm), small PEGDA discs ($r_p \sim 335$ µm, $Bo \sim 0.02$) with density $\rho_p \sim 1100$ kg/m$^3$ are predicted to remain in the quadrupolar regime ($\alpha \approx \frac{1}{6}$; Fig. 4b), while increasing particle density shifts $\alpha$ toward $\frac{1}{2}$ as negative-charge monopolar curvature strengthens with gravitational loading (Fig. 4d). In contrast, TMPTA–TMPTA pairs, which carry positive capillary charges at the same fluidic interface, exhibit the opposite trend: increasing $\rho_p$ suppresses the monopolar amplitude, driving $\alpha$ downward from $\frac{1}{2}$ toward $\frac{1}{6}$, signaling a resurgence of quadrupolar characteristics (Fig. 4f). The dichotomy arises because gravitational loading generally increases the monopolar amplitude $H_0$ for negative charged hydrophilic particles (Fig. 4d) [25], whereas for hydrophobic particles with positive capillary charges, gravity dampens the monopolar deformation (Fig. 4h) [51,52]. Equivalent parameter sweeps for Case 1 are provided in the SI (Figure S7).

*Contact author: atliu@umich.edu

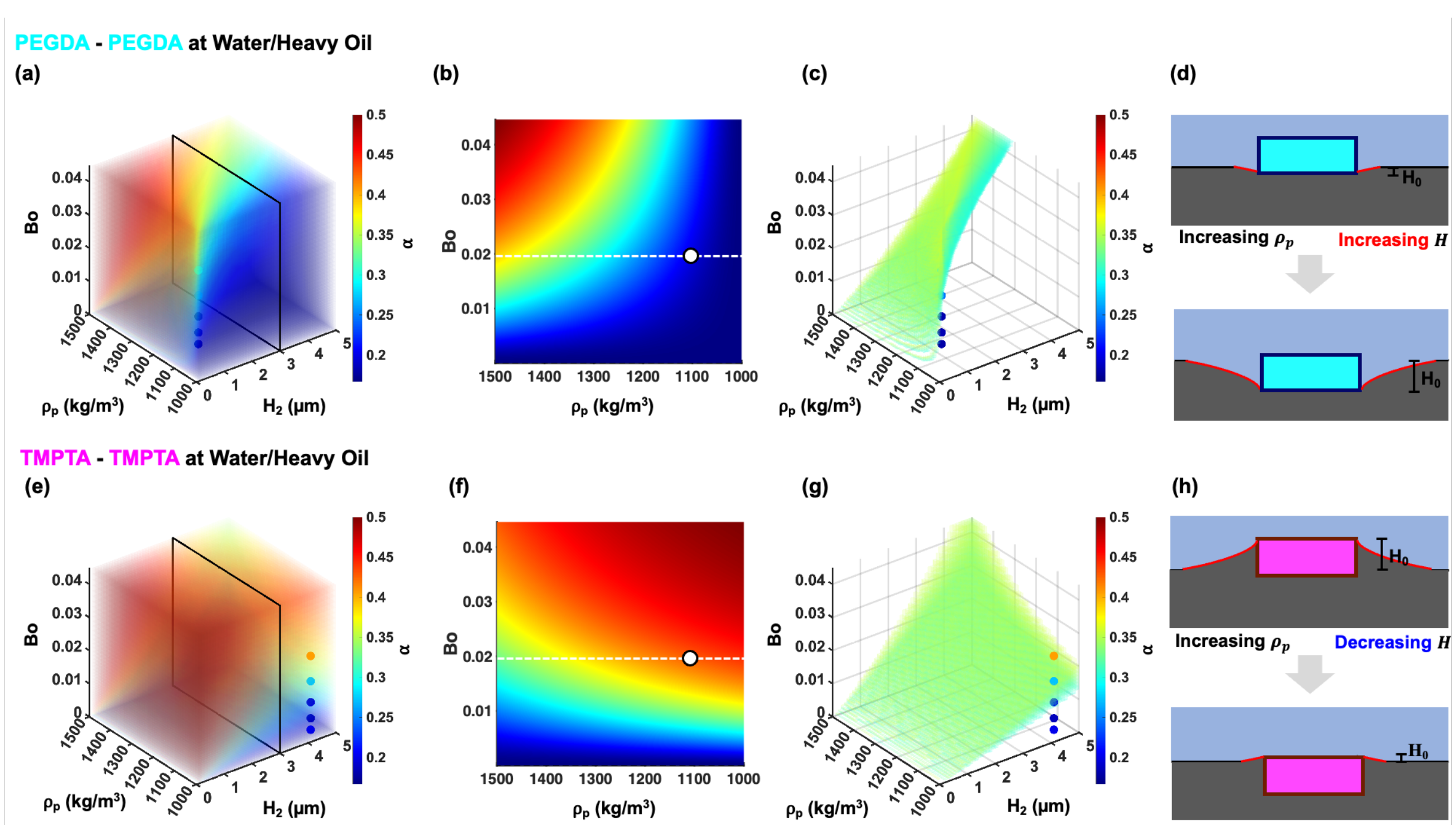

**Figure 4. Three-dimensional (3D) model prediction of the trajectory exponent $\alpha$ for capillary interactions between micro-disc particle pairs.** (a, e) Model predicted $\alpha$ values for (a) PEGDA–PEGDA and (e) TMPTA–TMPTA micro-disc pairs at the water/heavy oil interface. Each 3D surface plot maps $\alpha$ as a function of three input parameters: particle radius (via $Bo$), particle density ($\rho_p$), and the quadrupolar amplitude ($H_2$). Color maps correspond to $\alpha$ values, ranging from $\frac{1}{6}$ (quadrupolar regime) to $\frac{1}{2}$ (monopolar regime). (b, f) The central 2D plots isolate the $\rho_p$–$Bo$ plane at a fixed quadrupolar amplitude ($H_2$ = 3 μm). (c, g) 3D plots isolate the transition region defined as 0.30 < $\alpha$ < 0.36, highlighting the crossover between capillary regimes. Color-coded experimental $\alpha$ values (from Figure 3) are overlaid for comparison. (d, h) Schematic illustrations of how monopolar amplitude ($H_0$) changes with increasing particle density at the water/heavy oil interface: (d) for PEGDA particles with negative capillary charges, $H_0$ increases with density due to enhanced downward gravitational deformation; (h) for TMPTA particles with positive capillary charges, increasing density results in suppressed monopolar amplitude.

Isolating $\alpha$ near the critical exponent $\alpha^{\text{crit}} = 1/3$ (midpoint between 1/6 and 1/2) reveals that the transition is not set by $Bo$ alone, but by a coupled interplay of radius, density, and $H_2$ (Fig. 4c,g). This multivariate dependence motivates a closed-form criterion incorporating all three parameters. To accomplish this, we impose an overdamped force balance at the critical transition exponent $\alpha^{\text{crit}} = 1/3$ giving:

$$F_{drag}^{\text{crit}} \sim -\mathrm{C_D L^{-2}} \tag{7}$$

where $C_D$ is an as-yet-undetermined function of other relevant physical parameters. Details of this derivation are presented in the Appendix B1. We then equate $F_{\text{drag}}^{\text{crit}} = F_{cap} = F_{cap,0} + F_{cap,2}$ , where $F_{cap,0}$ and $F_{cap,2}$ are the monopolar and quadrupolar contributions given by Eq. 5–6, respectively. This balance yields an implicit transition condition in terms of $r_p$, $\rho_p$, $H_2$, $L$, and $C_D$.

Because a single exponent $\alpha$ characterizes the entire trajectory over the separations used to fit the power law, we evaluate the transition at an intermediate separation, setting $L = 4r_p$, the midpoint of the interparticle separation range used in simulations, which provides a consistent reference without introducing additional fitting degrees of freedom. For example, for PEGDA-PEGDA discs at the light oil/water interface, substitution of the force-balance solution for the monopolar amplitude into Eq. 5–6 yields the following closed-form capillary force:

$$F_{cap} = -2\pi\gamma \frac{\left(\frac{-\pi r_p^2 T g\rho_I + \pi r_p^2 T g \rho_p}{\frac{2\pi r_p \gamma}{r_p \ln\left(\frac{2\lambda_c}{\gamma r_p}\right)} + \pi r_p^2 g \rho_I}\right)^2}{4r_p} - 48\pi\gamma H_{2,1} H_{2,2} \frac{r_p^4}{\left(4r_p\right)^5} \quad (8)$$

where $H_{0,i}$ are determined from the coupled meniscus relations (Eq. 3–4) for the appropriate particle/interface configuration (full expressions for all cases are provided in the SI, Eq. S9–S11). The remaining task is to identify an appropriate expression for $C_D$ that enables a closed-form transition surface description in terms of experimentally accessible parameters $r_p$, $\rho_p$, and $H_2$.

*Contact author: atliu@umich.edu

Both a Stokes-like drag coefficient (Fig. 4a and S8) and a refinement accounting for viscosity partitioning between phases (Fig. 5a and S9) failed to reproduce the simulated transition. These models and their deviations are detailed in the Appendix B.

Guided by these observations, we incorporated the increased interfacial area ($S$) associated with the quadrupolar deformation into the effective drag. Larger roughness-induced quadrupolar amplitudes $H_2$ enlarge the distorted interfacial region (Fig. 5b), increasing viscous dissipation during lateral motion. The interfacial height profile ($h$) associated with the quadrupole deformation in polar coordinates is well-established [13,19,38,43]:

$$h = H_2 cos(2\varphi)\left(\frac{r_p}{r}\right)^2 \quad (9)$$

and can be used to estimate the roughness-induced excess interfacial area:

$$S = 2\pi \int_{r_p}^{\lim_{h\to 0} r} r\sqrt{1 + \frac{4H_2^2 r_p^4}{r^6}}\, dr \quad (10)$$

which is approximately linear in $H_2$ and weakly dependent on $r_p$ (Fig. S10). This motivates a simplified drag coefficient:

$$C_D(r_p, H_2) = \Lambda(1 + \zeta H_2) r_p \quad (11)$$

with global fitting parameters $\Lambda, \zeta$. Substituting Eq. 11 into Eq. 7 yields transition surfaces that closely match simulations across all pair/interface combinations (Fig. 5c,e–h), in contrast to radius-only or immersion-depth-dependent models.

Quantitative comparison via root-mean-square error (RMSE) relative to the critical exponent $\alpha^{crit} = 1/3$ (Appendix B4) confirms this improvement (Fig. 5d). The radius-only model ($C_D(r_p)$) gives an average RMSE of 0.917; including density ( $C_D(r_p, \rho_p)$ ) reduces error marginally to 0.899 (<2%). In contrast, $C_D(r_p, H_2)$ lowers RMSE by 28.3%, underscoring that quadrupolar amplitude (contact-line undulation) is essential for a predictive transition criterion.

Our framework provides a physically grounded description of capillary interactions that outperforms traditional $Bo$ -based scaling. The model also enables inverse inference of quadrupolar amplitude $H_2$ from trajectory data—a key parameter otherwise difficult to measure directly.

Knowing that monopolar and quadrupolar capillary interactions obey different force laws and approach kinetics, we hypothesize that collective self-assembly at fixed interface can be programmed using microparticles of identical chemical composition simply by tuning particle size. Our model predicts that, the transition between capillary interaction regimes should occur between radii of 250 µm and 375 µm at the water/heavy oil interface for both PEGDA and TMPTA micro-discs (Fig. 2h–j and 3b). To explore how this transition influences many-body interfacial self-organization, we dispersed mixed populations of micro-discs (equal number of PEGDA and TMPTA) at these two radii and monitored their assembly dynamics over one hour.

To quantify the evolving mesoscale organization, we analyzed radial particle distributions relative to the cluster centroid using kernel density estimation (KDE). Figure 6c–d present normalized radial distance KDE profiles for the large ($r_p$ = 375 µm) and small ($r_p$ = 250 µm) particle populations, respectively. In the monopolar dominant regime (Fig. 6c), PEGDA (cyan) and TMPTA (magenta) distribution initially overlap (t = 1 min) but separate progressively by t = 60 min: TMPTA particles concentrate near the cluster core while PEGDA particles shift toward the periphery, indicating wettability-driven sorting and phase demixing. In the quadrupolar regime (Fig. 6d), both distributions migrate toward the centroid and remain largely coincident, consistent with cooperative clustering and compositionally mixed aggregates. These trends were confirmed across three experimental replicates (Fig. S11). Tracking the KDE peak positions over time (Fig. 6e and 6f) confirms a clear bifurcation: particles with $r_p$ above the transition threshold undergo phase separation, whereas particles below it remain phase mixed. Time-lapse videos of both regimes, along with details on KDE construction and representative histograms, are provided in the SI (Movies S5–S6, Figure S12).

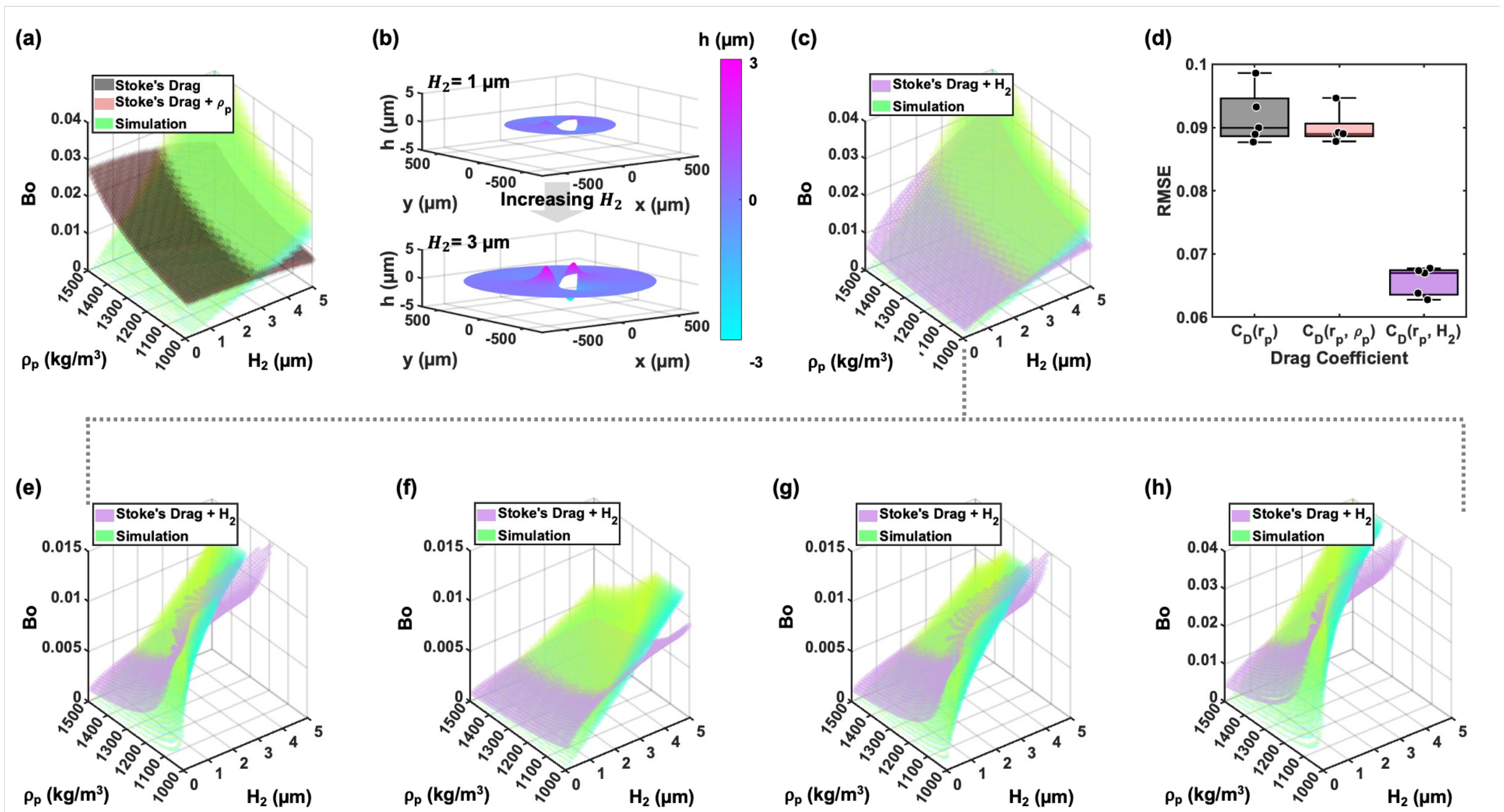

**Figure 5. Derivation of a closed-form criterion for capillary-regime transition.** (a) Predicted transition exponent $\alpha$ for TMPTA–TMPTA particle pairs at the water/heavy oil interface as a function of Bond number ($Bo$, representing particle radius), particle density ($\rho_p$), and quadrupolar amplitude ($H_2$). Surfaces are computed from the analytical model using (i) a traditional Stokes-like drag coefficient $C_D(r_p)$ and (ii) density-dependent drag coefficient $C_D(r_p, \rho_p)$. The translucent point cloud denotes the simulation-derived transition region (0.30 < $\alpha$ < 0.36). (b) Representative quadrupolar interfacial distortion profiles for a micro-disc ($r_p$=125 μm) with quadrupolar amplitudes $H_2$ = 1 μm and $H_2$ = 3 μm, illustrating that larger $H_2$ expands the deformed interfacial area (maximum radius defined by $h = 100$ nm). (c) Transition surface predicted using the proposed drag model $C_D(r_p, H_2)$, which incorporates the quadrupolar interfacial deformation area. This model shows strong agreement with the simulated transition surface reflecting ground truth. (d) Quantitative comparison of the three drag models via the root-mean-square error (RMSE) of the predicted $\alpha$ values. The proposed model incorporating the quadrupolar amplitude ($C_D(r_p, H_2)$) yields the lowest error across systems. (e-h) Transition surfaces or the remaining particle pair and interface combinations using $C_D(r_p, H_2)$: (e) PEGDA–PEGDA, (f) TMPTA–TMPTA, and (g) PEGDA–TMPTA at the light oil/water interface, and (h) PEGDA–PEGDA at the water/heavy oil interface.

This dichotomy in assembly behavior stems from the competition between unlike-pair monopolar repulsion (Fig. 6a) and quadrupolar attraction (Fig. 6b), which can be precisely modulated by adjusting particle size. We demonstrate, for the first time that inherently repulsive monopolar interactions, arising from mismatched wettability, can be overcome by size reduction alone, which suppresses gravitational deformation and reinstates quadrupolar attraction. Collectively, these findings establish particle size, density and surface topography (i.e., quadrupole amplitude) as practical design knobs for programmable two-dimensional particle self-assembly at fluidic interfaces.

## III. CONCLUSIONS

In this study, we systematically investigated the capillary interactions of disc-shaped microparticles at two model fluidic interfaces, light oil/water and water/heavy oil, across a range of particle sizes and wettabilities. Our experiments revealed a clear transition between monopole–monopole and quadrupole–quadrupole capillary interaction regimes within a compositionally invariant particle system. This transition was quantitatively characterized using the power-law trajectory exponents $\alpha$, which served as a robust indicator of the dominant capillary interaction mode. Complementing our experimental observations, we developed a predictive analytical model incorporating particle radius, density, and surface topography, and validated it against trajectory data across particle types and interfaces. The model exhibited excellent agreement with experimental data, accurately capturing the transition point and resolving the interplay between gravitational and topographic contributions to interfacial deformation. Finally, we demonstrate size-programmable many-body assembly at the water/heavy oil interface: large discs phase-separate via like–like attraction and unlike repulsion in the monopolar regime, whereas smaller discs form compositionally mixed clusters governed by quadrupolar attraction. Together, our empirical and theoretical findings elucidate a multivariate parameter space under which lateral capillary forces shift from gravitationally dominated monopolar interaction to topography-driven quadrupolar attraction. This unified framework offers critical insights into the interfacial assembly behavior of microscale particle systems and establishes practical design rules for tuning capillary interactions via geometric and topographic engineering.

*Contact author: atliu@umich.edu

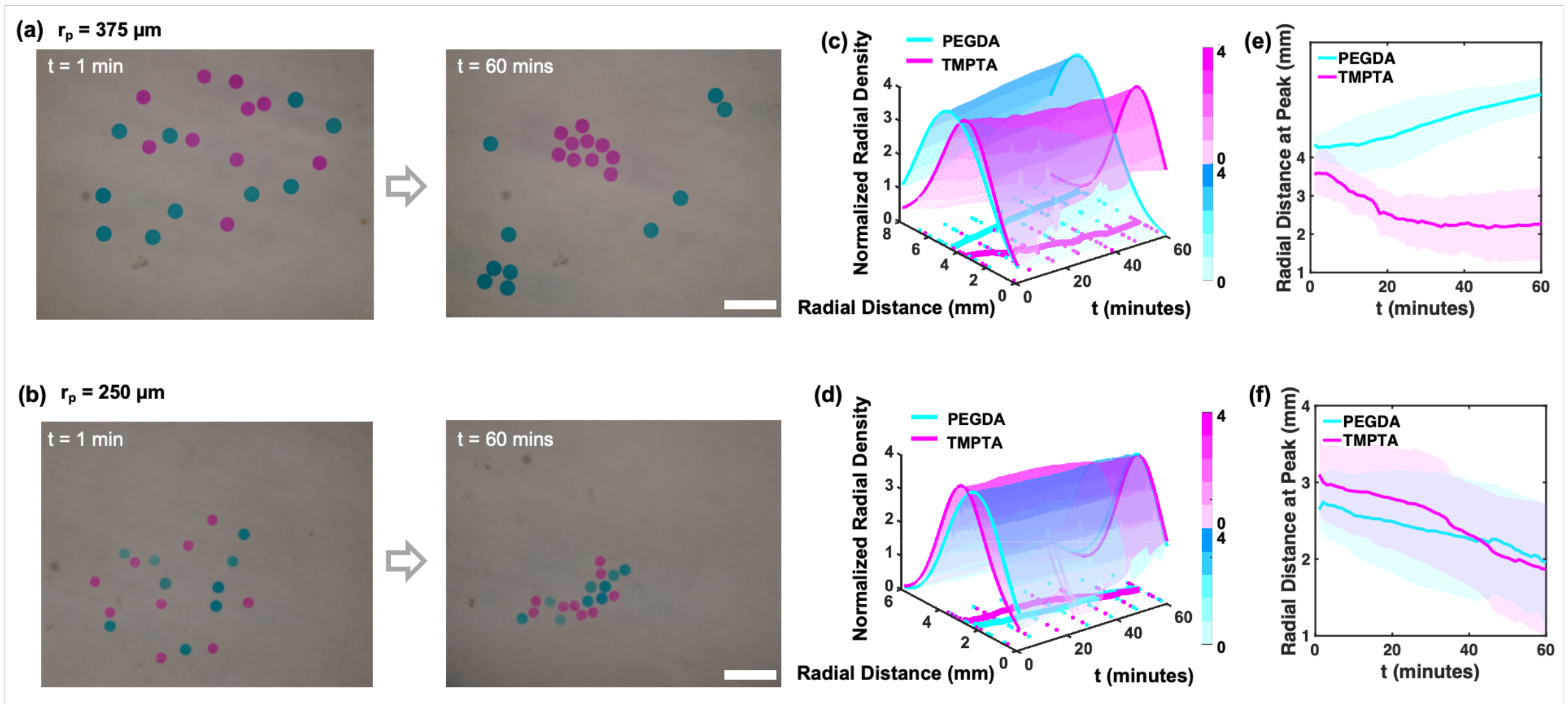

**Figure 6. Size-dependent self-assembly of micro-discs at the water/heavy oil interface.** (a) Monopolar regime: a mixed population of 10 PEGDA and 10 TMPTA micro-discs ($r_p$ = 375 µm) undergoes collective phase separation, driven by like-pair monopolar attraction and unlike-pair repulsion. (b) Quadrupolar regime: a mixed population of 9 PEGDA and 9 TMPTA micro-discs ($r_p$ = 250 µm) coalesce into a single, compositionally mixed cluster, consistent with quadrupolar capillary attraction. Scale bars: 3 mm. (c–d) Time-resolved radial distributions relative to the cluster centroid, quantified by 3D kernel density estimates (KDEs), for (c) $r_p$ = 375 µm and (d) $r_p$ = 250 µm. Individual particle positions (dots) and KDE ridgelines (curves) are overlaid in distance–time space for PEGDA (cyan) and TMPTA (magenta). (e–f) Sample-averaged KDE peak positions (n = 3) show (e) progressive phase separation in the monopolar regime and (f) sustained phase mixing in the quadrupolar regime. Scale bars: 3 mm.

## ACKNOWLEDGMENTS

This work is supported by the National Science Foundation under Award No. 2243104. The work was also supported by ACS PRF Doctoral New Investigator Award (66979-DNI10). The authors are thankful to Professor Anish Tuteja's group for usage of the 3D optical profilometer. The authors are also thankful to Professor Abdon Pena-Francesch's group for usage of the Dino-Lite lateral camera. The authors wish to acknowledge the support from the University of Michigan College of Engineering and the Michigan Center for Materials Characterization via an NSF grant (DMR-1625671), which supports the SEM resources that were utilized in this work.

*Contact author: atliu@umich.edu

*Contact author: atliu@umich.edu

*Contact author: atliu@umich.edu

## APPENDIX A

### A1. *Materials*

Silicon wafers (100 mm) were obtained from Silicon Valley Microelectronics (Santa Clara, CA). SU-8 2100 photoresist and developer were purchased from Kayaku Advanced Materials (Westborough, MA). Polyethylene glycol diacrylate (PEGDA, MW = 575 Da), pentaerythritol tetracrylate, trimethylolpropane triacrylate (TMPTA), lauryl acrylate, rhodamine B, and 2-hydroxy-2-methylpropiophenone (Darocur 1173) were sourced from Sigma-Aldrich (St. Louis, MO). Polydimethylsiloxane (PDMS; SYLGARD® 184 Silicone Elastomer Kit) was purchased from Dow Corning (Midland, MI). All reagents were used as received without further purification.

### A2. *Micro-disc Particles Fabrication*

Polymer micro-discs were fabricated via degassed micromolding lithography (Figure S1b), adapted from prior protocols [39]. In brief, SU-8 2100 photoresist was spin-coated onto silicon wafers and patterned by UV photolithography to generate master molds (Figure S1a) [53]. PDMS (10:1 base to curing agent) was cast on the SU-8 masters and cured at 60 °C for 5 hours to yield negative molds. Both molds and PDMS slabs were degassed at 0.05 Torr for 1 hour prior to use. Two photocurable precursor formulations were prepared. Hydrophilic (PEGDA): 700 µl PEGDA, 300 µl pentaerythritol tetracrylate, 50 µl Darocur 1173, and blue food dye. Hydrophobic (TMPTA): 900 µl TMPTA, 100 µl lauryl acrylate, 50 µl Darocur 1173, and rhodamine B [27]. Each precursor was dispensed onto a PDMS slab, followed by placement of the degassed PDMS mold (feature side down). After capillary filling, the assembly was gently compressed to remove excess precursor and exposed to 365 nm UV light for 1 minute (OmniCure LX500). Cured micro-discs were released by swelling the PDMS mold in dichloromethane, then rinsed with ethanol five times. PEGDA discs were additionally rinsed with Milli-Q water five times. To preserve wettability, PEGDA and TMPTA particles were stored in water and ethanol, respectively. Representative bright-field micrographs of the rinsed micro-discs with varying radii are provided in Figure S1c.

### A3. *Capillary Interaction Measurements*

Pairwise capillary interactions were measured at light oil/water and water/heavy oil interfaces in a 150 mL crystallization dish. The light oil phase consisted of a mixture of decane (70.9% vol.) and undecane (29.5% vol.), and Perfluorodecalin (PFD) was used as the heavy oil. The aqueous phase was Milli-Q water (18.2 MΩ·cm, TOC 1.7 ppb). After washing to remove solvent residues, micro-discs dispersed in water were deposited at the interface and allowed to equilibrate. Pairwise interactions were recorded at 30 fps using a digital single-lens reflex camera (SONY $\alpha$ 7R). Particle center-to-center separations were extracted from videos using VideoTracker.

### A4. *Water Contact Angle Measurements under Oil*

To quantify wettability, PEGDA and TMPTA precursor solutions were spin-coated onto glass substrates (2900 rpm, 30 seconds) and UV-cured for 2 minutes. Three-phase water contact angles were measured under each oil phase using a goniometer (Theta Lite, Biolin-Scientific). For water-in-heavy oil measurements, the cured films were inverted, and submerged in PFD, and 4 µl water droplets were deposited onto the surface. Contact angles were determined via Attension software (Biolin-Scientific).

### A5. *Quadrupolar Amplitude Measurements*

Interfacial deformation profiles were quantified by optical profilometry (LEXT OLS 5100 Laser Microscope, Evident Scientific). Micro-discs ($r_p$ = 125um) were deposited at light oil/water and water/heavy oil interfaces prepared in glass petri dishes. White-light interferometric scans were acquired using a 5× objective. Surface height maps were analyzed to extract quadrupolar amplitudes for individual particles at each interface.

### A6. *Lateral Imaging*

High-resolution side-view images of particles adsorbed at interfaces were acquired in quartz glass cells (DataPhysics, NC). PEGDA and TMPTA disc particles were deposited at light oil/water and water/heavy oil interfaces and imaged laterally using a Dino-Lite Edge Digital Microscope.

## APPENDIX B: Drag coefficient

*Contact author: atliu@umich.edu

**B1. Critical drag scaling from force balance**

Neglecting inertial and stochastic forces, lateral particle motion at a fluid interface is governed by an overdamped balance between capillary driving force and hydrodynamic resistance:

$$F_{\text{cap}} = F_{\text{drag}} = C_D\, v \quad \text{(B1)}$$

where $C_D$ is an effective drag coefficient, which is dependent on particle geometry, immersion depth, and fluid properties, and $v$ is the lateral velocity [53].

From the power-law trajectory relation (Eq. 2, main text), $L \sim (t_f - t)^{\alpha}$, the velocity scales as $v = dL/dt \sim -\alpha(t_f - t)^{\alpha-1}$. At the critical transition, $\alpha = \alpha^{\text{crit}} = 1/3$, giving $v \sim -(t_f - t)^{-2/3}$. Eliminating time via $t_f - t \sim L^{1/\alpha} \sim L^3$ yields $v \sim -L^{-2}$, and thus the critical drag scaling presented in Eq. 7 of the main text.

**B2. Baseline model: Radius-dependent drag**

As a baseline, we adopted a Stokes-like drag coefficient:

$$C_D(r_p) = \Lambda r_p \quad \text{(B2)}$$

where $\Lambda$ is a geometry-dependent prefactor [9]. Substituting this into Eq. 7 together with the capillary force expression (Eq. 8) produced a transition surface that deviates substantially from simulation results (Fig. 5a and S8). This failure indicates that particle radius alone cannot capture the quadrupolar–monopolar crossover.

**B3. Refinement: Immersion-depth-dependent drag**

We next incorporated viscosity partitioning between the two fluid phases by allowing $C_D$ to depend on immersion depth, which is set by the monopolar amplitude $H_0$ and therefore by particle density $\rho_p$ (Fig. 4d,h). For a TMPTA disc at the water/heavy oil interface, a thickness $T - H_0$ resides in the heavy oil subphase, while $H_0$ is exposed to the water superphase. The resulting two-layer viscous drag is:

$$C_D(r_p, \rho_p) = 2[\Lambda_1 \eta_I (T - H_0) + \Lambda_2 \eta_{II} H_0] r_p \quad \text{(B3)}$$

where $\eta_I$ and $\eta_{II}$ are the subphase and superphase viscosities, and $\Lambda_1, \Lambda_2$ are fitting constants. The prefactor of 2 accounts for unlike pairs in which the two interacting particles experience different immersion depths. Analogous expressions for other particle/interface combinations are provided in the SI (Eqs. S12–S14).

Despite this refinement, the predicted transition surface still failed to align with simulations (Fig. 5a), and similar discrepancies persisted across other cases (Fig. S9). Immersion depth alone was insufficient to capture the dissipation associated with quadrupolar interfacial deformations.

**B4. Quantitative model comparison**

To compare model performance, we computed the root-mean-square error (RMSE) between trajectory exponents $\alpha$ predicted by each closed-form drag model and those obtained from full numerical simulations. RMSE was evaluated over all sampled parameter sets ($r_p, \rho_p, H_2$) relative to the critical transition exponent $\alpha^{\text{crit}} = 1/3$:

$$\text{RMSE} = \sqrt{\frac{1}{n}\sum_{i=1}^{n}\left(\frac{1}{3} - \alpha_i\right)^2} \quad \text{(B4)}$$

The radius-only model ($C_D(r_p)$) yielded an average RMSE of 0.917; the immersion-depth model ($C_D(r_p, \rho_p)$) reduced this marginally to 0.899 (<2% improvement). These results are shown in Fig. 5d and Fig. S9 of the main text.